\documentstyle[prd,aps]{revtex}

\newcommand{\bea}{\begin{eqnarray}}
\newcommand{\eea}{\end{eqnarray}}

\begin{document}

\draft
\twocolumn[\hsize\textwidth\columnwidth\hsize\csname
@twocolumnfalse\endcsname

\title{Cosmological Perturbations with Multiple Scalar Fields}
\author{Jai-chan Hwang}
\address{Department of Astronomy and Atmospheric Sciences,
         Kyungpook National University, Taegu, Korea}
\date{\today}
\maketitle

\begin{abstract}

In this {\it brief note} we present a set of equations describing the 
evolution of perturbed scalar fields in a cosmological spacetime 
with multiple scalar fields. 
We take into account of the simultaneously excited full metric 
perturbations in the context of the uniform-curvature gauge 
which is known to be the best choice.
The equations presented in a compact form will be useful for handling 
the structure formation processes under the multiple episodes of inflation.

\end{abstract}


\vskip2pc]
Doule inflation recently attracted much attention in the literature
because it allows more freedom in designing the spectrums of the 
large scale structures in the generation stage.
There have been many attempts to calculate the generated large scale 
spectrums with multiple episodes of inflation
\cite{double-infl-other-gauge,double-infl-no-metric}.
In this note we will address an important point needed when we properly 
handle the quantum generation and classical evolution processes of the 
structures in relativistic gravity, namely, the gauge issue.

Past development in the field shows that the proper gauge choice 
is essential for properly handling the cosmological perturbations 
in relativistic gravity.
Under the proper gauge condition the analysis including the 
metric perturbations allows more general and consistent result
compared with the one ignoring them; examples are the case of the
minimally coupled scalar field \cite{H-QFT} and the generalized
gravity theories \cite{H-GGT}.
{}From a thorough study made in various gauge conditions 
we found that the uniform-curvature gauge choice fits the problems 
involving the scalar field \cite{H-MSF}.
However, {\it all} previous works on estimating the perturbation 
spectrum generated from multiple inflation stages are based on 
either other gauge \cite{double-infl-other-gauge} or 
ignoring the perturbed metric contributions \cite{double-infl-no-metric}.
As a matter of fact, similar situation prevails even in the case of the 
single scalar field.
Since the proper gauge choice is known in the literature,
it is not economic to continue work in other gauge conditions.
{}Furthermore, under the proper gauge condition, the equation including 
the perturbed metric looks no more complicated than ignoring it.
{}For the benefit of future studies, in this note 
we would like to present the perturbed scalar field equations 
valid for the multiple minimally coupled scalar fields
in the uniform-curvature gauge. 
The result is in Eq. (\ref{UCG-delta-phi-i-eq}).

We consider Einstein gravity with multiple minimally coupled 
scalar fields with a general potential
\bea
   L = {1 \over 16 \pi G} R - \sum_{k = 1}^{n} {1\over 2} 
       {\phi_{(k)}}^{;a} \phi_{(k),a} - V(\phi_{(j)}),
   \label{Lagrangian}
\eea
where $\phi_{(i)}$'s are the scalar fields 
and $V$ is a general function of $\phi_{(i)}$'s;
$i,j,k = 1 \dots n$, with $n$ arbitrary positive integer.

As a cosmological model describing the perturbed universe,
we consider a homogeneous and isotropic (flat) background with
the general scalar type perturbations
\cite{COMMENT-metric}
\bea
   d s^2
   &=& - \left( 1 + 2 \alpha \right) d t^2 - a \chi_{,\alpha} d t d x^\alpha
   \nonumber \\
   & & + \;
       a^2 \delta_{\alpha\beta} \left( 1 + 2 \varphi \right)
       d x^\alpha d x^\beta,
   \label{metric-general}
\eea
where $\alpha ({\bf x}, t)$, $\chi ({\bf x}, t)$, and $\varphi ({\bf x}, t)$
are the scalar type metric perturbations.
We consider perturbations in the scalar fields as
\bea
   & & \phi_{(i)} ({\bf x}, t) = \bar \phi_{(i)} (t) 
       + \delta \phi_{(i)} ({\bf x}, t),
\eea
where a background quantity is indicated by an overbar
which will be neglected unless necessary.
The equations describing the background universe are presented in 
Eqs. (21,65) and below Eq.(88) of \cite{H-PRW}:
\bea
   & & H^2 = {8 \pi G \over 3} 
       \left( {1 \over 2} \sum_{k=1}^{n} \dot \phi_{(k)}^2 + V \right),
   \label{BG1} \\
   & & \ddot \phi_{(i)} + 3 H \dot \phi_{(i)} 
       + V_{, \phi_{(i)}} = 0,
   \label{BG3}
\eea
where $H \equiv \dot a / a$.

The equations describing the perturbed part of the scalar fields and
the metric are presented in Eqs.(88,89) of \cite{H-PRW} without choosing
the temporal gauge condition, i.e., in a gauge ready form
\cite{COMMENT-gauge}.
We introduce gauge invariant combinations
\bea
   & & \delta \phi_{(i)\varphi}
       \equiv \delta \phi_{(i)} - {\dot \phi_{(i)} \over H} \varphi,
   \label{UCG-UFG}
\eea
which is $\delta \phi_{(i)}$ in the uniform-curvature gauge
which takes $\varphi = 0$ as the gauge condition.
Under this gauge condition, the perturbed part equations involving the
metric and scalar fields can be organized in the following compact forms
\bea
   & & \delta \ddot \phi_{(i)\varphi} + 3 H \delta \dot \phi_{(i)\varphi}
       - {1 \over a^2} \nabla^2 \delta \phi_{(i)\varphi}
   \nonumber \\
   & & \qquad
       + \sum_{k=1}^{n} \Bigg[ V_{, \phi_{(i)} \phi_{(k)} }
       - {8 \pi G \over a^3} \left( {a^3 \over H}
       \dot \phi_{(i)} \dot \phi_{(k)} \right)^\cdot
       \Bigg] \delta \phi_{(k)\varphi} = 0.
   \nonumber \\
   \label{UCG-delta-phi-i-eq}
\eea
This set of equations is the main result of this note.
These are coupled second order differential equations for
$\delta \phi_{(i)\varphi}$.
The coefficients follow the background equations in 
Eqs. (\ref{BG1},\ref{BG3}).
Thus, if we have a multiple inflation scenario satisfying Eqs. 
(\ref{BG1},\ref{BG3}) the evolution of the perturbations can be 
followed by using Eq. (\ref{UCG-delta-phi-i-eq}).
The second term in the bracket is originated from considering the 
perturbed metric contribution, see Eq. (88) of \cite{H-PRW}.
We note two points:
{}First, the perturbed metric contribution does not look complicated, 
and second, we believe that properly taking into account of the term 
will make further analyses self-consistent and rewarding.
In order to illustrate the second point we summarize the result
for a single component.

In the case of a single component Eq. (\ref{UCG-delta-phi-i-eq}) becomes
\bea
   & & \delta \ddot \phi_\varphi + 3 H \delta \dot \phi_\varphi
       + \Bigg[ - {1 \over a^2} \nabla^2 
   \nonumber \\
   & & \qquad
       + \; V_{, \phi \phi} 
       + 2 {\dot H \over H} \left( 3 H - {\dot H \over H}
       + 2 {\ddot \phi \over \dot \phi} \right) \Bigg] \delta \phi_\varphi = 0.
   \label{UCG-delta-phi-eq}
\eea
This equation was derived in \cite{H-QFT} and has unexpected 
interesting consequences:
{}First, it has general asymptotic solutions.
{}For example, in the large scale limit we have
\bea
   & & \delta \phi_\varphi ({\bf x}, t) = {\dot \phi \over H}
       \left[ - C ({\bf x})
       + D ({\bf x}) \int^t {H^2 \over a^3 \dot \phi^2} dt \right],
   \label{UCG-delta-phi-sol}
\eea
which is valid for the general $V(\phi)$.
We note that such a general solution is {\it not} available if we ignore 
the metric contributions.
Second, using Eq. (\ref{UCG-delta-phi-eq}) the quantum fluctuations 
of $\delta \hat \phi$ can be derived in {\it exact} analytic forms 
when the background scale factor follows the exponential or power-law type 
inflationary behaviors; see \cite{H-QFT}.

The recently popular extended inflation scenario necessarily involves
an additional inflation field.
There were many attempts to calculate generated density spectrum in the
scenario \cite{Extended-infl-pert}.
However, none of the work recognized the role of the uniform-curvature gauge,
and furthermore, all works in \cite{Extended-infl-pert}
are based on the conformal transformation technique the validity of which 
is dubious especially for handling the quantum generation processes. 
A thorough study in various gauge conditions in \cite{HN-GGT} shows that the 
uniform curvature gauge again suits for handling the generalized 
gravity theories. 
Perturbed set of equations in the case of generalized gravity with multiple 
minimally coupled scalar fields is presented in Sec. 4.2 of \cite{H-PRW} 
in the gauge ready form; the equations are presented in the original frame
of generalized gravity. 

Many of the previous works in 
\cite{double-infl-other-gauge,double-infl-no-metric,Extended-infl-pert} 
may deserve another look with a new perspective of the proper gauge.



\begin{references}
\bibitem{double-infl-other-gauge}
         A. A. Starobinsky, JETP Lett. {\bf 42}, 152 (1985);
         L. A. Kofman and D. Yu. Pogosyan, Phys. Lett. B {\bf 214}, 508 (1988);
         D. S. Salopek, J. R. Bond and J. M. Bardeen, Phys. Rev. D {\bf 40},
               1753 (1989);
         S. Mollerach, Phys. Rev. D {\bf 42}, 313 (1990);
         D. Polarski and A. A. Starobinsky, Nucl. Phys. B {\bf 385}, 623 (1992);
         D. Polarski, Phys. Rev. D {\bf 49}, 6319 (1994);
         P. Peter, D. Polarski and A. A. Starobinsky, Phys. Rev. D {\bf 50}, 
            4827 (1994);
         D. Polarski and A. A. Starobinsky, Phys. Rev. D {\bf 50}, 6123 (1994);
         D. Polarski and A. A. Starobinsky, Phys. Lett. B {\bf 356}, 196 (1995);
         A. A. Starobinsky and J. Yokoyama, gr-qc/9502002;
         J. Garc\'ia-Bellido and D. Wands, gr-qc/9506050;
         J. Yokoyama, gr-qc/9509006;
\bibitem{double-infl-no-metric}
         L. A. Kofman, A. D. Linde and A. A. Starobinsky, Phys. Lett. B 
               {\bf 157}, 361 (1985);
         A. D. Linde, Phys. Lett. B {\bf 158}, 375 (1985);
         L. A. Kofman and A. D. Linde, Nucl. Phys. B {\bf 282}, 555 (1987);
         J. Garc\'ia-Bellido, A. Linde and D. Wands, astro-ph/9605094;
         J. Garc\'ia-Bellido and D. Wands, astro-ph/9606047;
\bibitem{H-QFT}
         J. Hwang, Phys. Rev. D {\bf 48}, 3544 (1993);
         Class. Quant. Grav. {\bf 11}, 2305 (1994),
\bibitem{H-GGT}
         J. Hwang, Phys. Rev. D {\bf 53}, 762 (1996);
         gr-qc/9607059.
\bibitem{H-MSF}
         J. Hwang, Astrophys. J. {\bf 427}, 542 (1994).
\bibitem{COMMENT-metric}
         In the homogeneous and isotropic background the scalar, vector,
         and tensor type perturbations evolve independently of each other.
         The perturbations of scalar fields directly couple only with
         the scalar type perturbations. 
         In Eq. (\ref{metric-general}) we have taken a spatial gauge condition
         so that all the perturbed quantities are spatially gauge invariant.
\bibitem{H-PRW}
         J. Hwang, Astrophys. J. {\bf 375}, 443 (1991).
\bibitem{COMMENT-gauge}
         As a temporal gauge condition we can choose one
         among $\alpha =0$, $\chi = 0$, and $\varphi = 0$
         which correspond to the synchronous gauge, the zero-shear gauge,
         and the uniform-curvature gauge, respectively.
         There exist more fundamental gauge conditions available, 
         see Sec 3.3 of \cite{H-PRW}.
         We have an opinion on the gauge choice in a pragmatic sense: 
         depending on the problem, certain gauge condition is superior 
         in the sense that the calculation is simpler (or the analyses 
         can be done similarly to the case of Newtonian or ordinary quantum 
         field theory) than in other gauge conditions.
         Still, unless one makes mistakes, the final physically measurable 
         results should be the same independent of which gauge one has chosen.  
         In practice, we found certain calculation can be done only in 
         certain gauge and not in others which is the case of quantum field 
         analyses of $\delta \phi$, see below Eq. (\ref{UCG-delta-phi-eq}).
         Gauge ready method suggests that in case we do not know which gauge 
         condition is the best a priori, investigate all the gauge conditions.  
         We have gone through this process \cite{H-MSF} and ended up in the 
         uniform-curvature gauge for the scalar field.
\bibitem{Extended-infl-pert}
         A. H. Guth and B. Jain, Phys. Rev. D {\bf 45}, 426 (1992);
         S. Nakamura and A. Hosoya, Prog. Theor. Phys. {\bf 87}, 401 (1992);
         S. Mollerach and S. Matarrese, Phys. Rev. D {\bf 45}, 1961 (1992);
         N. Deruelle, C. Gundlach and D. Langlois, Phys. Rev. D {\bf 46}, 
            5337 (1992);
         A. M. Laycock and A. R. Liddle, Phys. Rev. D {\bf 49}, 1827 (1994);
         M. Kuwahara, H. Suzuki and E. Takasugi, Phys. Rev. D {\bf 50}, 661
            (1994);
         A. M. Green and A. R. Liddle, astro-ph/9604001, astro-ph/9607166.
\bibitem{HN-GGT}
         J. Hwang and H. Noh, Phys. Rev. D {\bf 54}, 1460 (1996).
\end{references}
\end{document}